\documentclass[12pt]{article}
\usepackage{graphicx}
\usepackage{amsmath}
\usepackage{amsfonts}
\usepackage{amssymb}
\newtheorem{theorem}{Theorem}

\newtheorem{remark}[theorem]{Remark}

\begin{document}

\title{Applications of the Wulff construction to the number theory}
\author{Senya Shlosman\thanks{The work was partially supported by the Russian Fund for
Fundamental research through grant 99-01-00284.}\\Centre de Physique Theorique, \\CNRS, Luminy, case 907, F-13288, \\Marseille Cedex 9, France, and \\IITP, RAS, Moscow, Russia\\\textit{shlosman@cpt.univ-mrs.fr}}
\maketitle
\begin{abstract}
We apply the geometric construction of solutions of some variational problems
of combinatorics to estimate the number of partitions and of plane partitions
of an integer.

\textbf{Key words and phrases:} Wulff construction, Young diagram, plane partition.
\end{abstract}

\section{Statement of results.}

In this note we will explain how certain classical statements from the number
theory can be rederived using recent results of the asymptotic combinatorics.
Specifically, we will derive the asymptotic behavior of the number of the
partitions and of the plain partitions of an integer $N.$

We recall that a partition $p$ of an integer $N$ is an array of non-negative
integers $n_{1}\geq n_{2}\geq...\geq n_{k}\geq...,$ such that $\sum
_{i=1}^{\infty}n_{i}=N.$ One corresponds to $p$ a well-known geometric object
called Young diagram, $Y\left(  p\right)  ,$ with $N$ cells. The set of all
Young diagrams with $N$ cells will be denoted by $\mathcal{Y}_{N}.$

Similarly, a plane partition $S$ of an integer $N$ is a two-dimensional array
of non-negative integers $n_{ij},$ such that for any $i$ we have $n_{i1}\geq
n_{i2}\geq...\geq n_{ik}\geq...,$ for any $j$ we have $n_{1j}\geq n_{2j}%
\geq...\geq n_{kj}\geq...,$ while again $\sum_{i,j=1}^{\infty}n_{ij}=N.$ The
corresponding geometric picture is called a \textit{skyscraper, }or a
\textit{mausoleum}, or a \textit{3D Young diagram}. The set of all skyscrapers
with $N$ cells ($=$ the set of all plane partitions of $N$) will be denoted by
$\mathcal{S}_{N}.$

Our goal is to explain that as $N\rightarrow\infty$
\begin{equation}
\ln\left|  \mathcal{Y}_{N}\right|  \sim\pi\left(  \frac{2}{3}\right)
^{1/2}N^{1/2}, \label{a1}%
\end{equation}
and
\begin{equation}
\ln\left|  \mathcal{S}_{N}\right|  \sim3\left(  \frac{\zeta\left(  3\right)
}{4}\right)  ^{1/3}N^{2/3}. \label{a2}%
\end{equation}
In fact, much more precise information about the behavior of these functions
is known. For $\left|  \mathcal{Y}_{N}\right|  $ it is the famous result of
Hardy-Ramanujan-Rademacher, see \cite{An}. The corresponding asymptotic
behavior of $\left|  \mathcal{S}_{N}\right|  $ was found by Wright in
\cite{Wr}. \medskip

To get the relations (\ref{a1}) and (\ref{a2}), we are using, correspondingly,
the results of \cite{VKer} and \cite{CKe}, where the shape of the typical
Young diagram, resp. skyscraper, were found. (Here ``typical'' means with
respect to the uniform distribution on the sets $\mathcal{Y}_{N}$ and
$\mathcal{S}_{N}.$) These shapes are solutions of certain variational
maximizing problems, described in details in the next section. Briefly
speaking, these problems consist in finding a hypersurface $G_{\eta}$ in a
certain class, which maximizes the surface integral $\mathcal{V}_{\eta}\left(
G\right)  =\int_{G}\eta\left(  \mathbf{n}_{x}\right)  \,ds_{x},$ where
$\mathbf{n}_{x}$ is the normal to $G$ at $x,$ and $\eta$ is a corresponding
given function.

For the case of Young diagrams the function $\eta_{\mathcal{Y}}$ is defined on
a unit quarter-circle $\Delta^{1}=S^{1}\cap\mathbb{R}_{+}^{2},$ and is given
by
\[
\eta_{\mathcal{Y}}\left(  \mathbf{n}\right)  =-\left(  n_{1}\ln\frac{n_{1}%
}{n_{1}+n_{2}}+n_{2}\ln\frac{n_{2}}{n_{1}+n_{2}}\right)  .
\]
The typical Young diagram with $N$ cells, scaled by the factor $N^{-1/2}$,
goes along the curve $\mathcal{C}:$%
\[
\exp\left\{  -\frac{\pi}{\sqrt{6}}x\right\}  +\exp\left\{  -\frac{\pi}%
{\sqrt{6}}y\right\}  =1.
\]
This result was obtained by Vershik, see \cite{VKer} or \cite{V}.

In the skyscraper case the function $\eta_{\mathcal{S}}$ is defined on the set
$\Delta^{2}=S^{2}\cap\mathbb{R}_{+}^{3}$ of unit vectors $\mathbf{n}$ and is
given by
\[
\eta_{\mathcal{S}}\left(  \mathbf{n}\right)  =\frac{\left|  \mathbf{n}\right|
_{1}}{\pi}\sum_{i=1}^{3}L\left(  \pi\frac{n_{i}}{\left|  \mathbf{n}\right|
_{1}}\right)  ,
\]
where $\left|  \mathbf{n}\right|  _{1}=n_{1}+n_{2}+n_{3},$ and $L$ is the
Lobachevsky function, defined for $x\in\left[  0,\pi\right]  $ by
\[
L\left(  x\right)  =-\int_{0}^{x}\ln\left(  2\sin t\right)  dt.
\]
The surface $G_{\eta_{\mathcal{S}}}$ in $\mathbb{R}^{3},$ which describes the
shape of a typical skyscraper with $N$ cells scaled by the factor $N^{-1/3}$
to the unit volume, is given by
\begin{align*}
&  G_{\eta_{\mathcal{S}}}=\\
&  =\left(  \frac{\zeta\left(  3\right)  }{4}\right)  ^{-1/3}\left\{  \left(
f\left(  A,B,C\right)  -\ln A,f\left(  A,B,C\right)  -\ln B,f\left(
A,B,C\right)  -\ln C\right)  \right\}  ,
\end{align*}
where $A,B,C>0,\;A+B+C=1,$ and%
\[
f\left(  A,B,C\right)  =\frac{1}{4\pi^{2}}\int_{\left[  0,2\pi\right]  }%
\int_{\left[  0,2\pi\right]  }\ln\left(  A+Be^{iu}+Ce^{iv}\right)  dudv,
\]
see \cite{CKe}.\medskip

The geometric construction, which gives the solution to problems described
above, was found in \cite{S1}. In fact, it is very close to the Wulff
construction, \cite{W}. Both constructions are described below, together with
the relation between them. Due to the relation (\ref{aa}), the evaluation of
the integrals $\mathcal{V}_{\eta}\left(  G_{\eta}\right)  $ becomes very easy.
It turns out that
\begin{equation}
\mathcal{V}_{\eta_{\mathcal{Y}}}\left(  G_{\eta_{\mathcal{Y}}}\right)
=\pi\left(  \frac{2}{3}\right)  ^{1/2}, \label{a3}%
\end{equation}
while
\begin{equation}
\mathcal{V}_{\eta_{\mathcal{S}}}\left(  G_{\eta_{\mathcal{S}}}\right)
=3\left(  \frac{\zeta\left(  3\right)  }{4}\right)  ^{1/3}. \label{a4}%
\end{equation}
We know from \cite{VKer} that most of the Young diagrams with $N$ cells go
along the curve $N^{1/2}\mathcal{C}.$ On the other hand, if $G$ is a graph of
a decaying function $g\left(  x\right)  $ with $\int_{0}^{\infty}g\left(
x\right)  dx=N,$ then the meaning of the functional $\mathcal{V}%
_{\eta_{\mathcal{Y}}}\left(  G\right)  $ is the following: ``the number of the
diagrams with $N$ cells going along the curve $G$ is of the order of
$\exp\left\{  \mathcal{V}_{\eta_{\mathcal{Y}}}\left(  G\right)  \right\}  $''.
Therefore, the value (\ref{a3}), multiplied by $N^{1/2}$ is precisely the
exponent of (\ref{a1}). Correspondingly, most of the skyscrapers with $N$
cells go along the surface $N^{1/3}G_{\eta_{\mathcal{S}}},$ see \cite{CKe}.
Therefore the value $\mathcal{V}_{\eta_{\mathcal{S}}}\left(  N^{1/3}%
G_{\eta_{\mathcal{S}}}\right)  =N^{2/3}$ $\mathcal{V}_{\eta_{\mathcal{S}}%
}\left(  G_{\eta_{\mathcal{S}}}\right)  \ $has to be the exponent (\ref{a2}).
For more details the reader should consult the papers \cite{VKer}, \cite{DVZ},
\cite{CKe} and the review paper \cite{S2}.

\section{The maximizing problem.}

Let $S^{d}$ denote the $d$-dimensional unit sphere in $\mathbb{R}^{d+1}.$
Introduce the subset $\Delta^{d}=S^{d}\cap\mathbb{R}_{+}^{d+1}\ $of
``positive'' unit vectors, lying in the positive octant. Let a function $\eta$
on $\Delta^{d}$ be given, which is supposed to be continuous, nonnegative:
$\eta\left(  \mathbf{\cdot}\right)  \geq0,$ and vanishing on $\partial
\Delta^{d}.$ We suppose for simplicity that the function $\eta$ vanishes also
in some neighborhood of $\partial\Delta^{d}.$ (The general result which is
needed for the formulas in the previous section is then obtained by letting
this neighborhood to shrink to zero.) Let now $G\subset\mathbb{R}^{d+1}$ be an
embedded hypersurface, possibly with a boundary. We assume that for almost
every $x\mathbf{\in}G$ the normal vector $\mathbf{n}_{x}$ to $G$ is defined,
and moreover
\begin{equation}
\mathbf{n}_{x}\in\Delta^{d}\text{ for a.e. }x\mathbf{\in}G. \label{07}%
\end{equation}
Then we can define the functional
\[
\mathcal{V}_{\eta}\left(  G\right)  =\int_{G}\eta\left(  \mathbf{n}%
_{x}\right)  \,ds_{x},
\]
here $ds$ is the usual volume $d$-form induced from the Riemannian metric on
$\mathbb{R}^{d+1}$ by the embedding $G\subset\mathbb{R}^{d+1}.$

Denote by $Q_{N}\subset\mathbb{R}_{+}^{d+1}$ the cube, consisting of points
$x\in\mathbb{R}^{d+1}$ with $0\leq x_{i}\leq N,$ $i=1,2,...,d+1.$ Denote by
$O$ the point $\left(  0,0,...,0\right)  \in Q_{N},$ and let $A=\left(
N,N,...,N\right)  \in Q_{N}$ be the opposite vertex.

Let the number $V\in\left(  0,N^{d+1}\right)  $ be given. We introduce the
family $D_{V}^{N}$ of hypersurfaces $G\subset Q_{N}$ as follows:

\noindent$\,i)$ $\,G$ splits the cube $Q_{N}$ into two parts, with the points
$O$ and $A$ belonging to different parts; denote them by $Q_{N}\left(
G,O\right)  $ and $Q_{N}\left(  G,A\right)  ;$

\noindent$ii)$ the property (\ref{07}) holds

\noindent$iii)$ for the volume $\mathrm{vol}\left(  G,N\right)  ,$ defined as
the ($\left(  d+1\right)  $-dimensional) volume $\mathrm{vol}\left(
Q_{N}\left(  G,O\right)  \right)  $ of the body $Q_{N}\left(  G,O\right)  ,$
we have
\[
\mathrm{vol}\left(  G,N\right)  =V.
\]

We want to solve the following variational problem: find the \textit{upper
bound} of $\mathcal{V}_{\eta}$ over $D_{V}^{N}:$%
\begin{equation}
v_{\eta}=\sup_{G\in D_{V}^{N}}\mathcal{V}_{\eta}\left(  G\right)  , \label{04}%
\end{equation}
as well as the maximizing surface(s) $V_{\eta}\in D_{V}^{N},$ such that
$\mathcal{V}_{\eta}\left(  V_{\eta}\right)  =v_{\eta},$ if the maximizer does exist.

It turns out that there exists a geometric construction, which provides a
solution to the variational problem (\ref{04}). It was found in \cite{S2}.

Let
\begin{equation}
K_{\eta}^{>}=\left\{  \mathbf{x}\in\mathbb{R}_{+}^{d+1}:\forall\mathbf{n}%
\in\Delta^{d}\;\left(  \mathbf{x},\mathbf{n}\right)  \geq\,\eta\left(
\mathbf{n}\right)  \right\}  , \label{15}%
\end{equation}%
\begin{equation}
G_{\eta}=\partial\left(  K_{\eta}^{>}\right)  . \label{16}%
\end{equation}
Note that $K_{\eta}^{>}$ is a convex set, so the hypersurface $G_{\eta}$ has
normals at almost every point. Moreover, these normals fall into $\Delta^{d}.$
Define the dilatation parameter $\lambda\left(  V,N\right)  $ as a unique
solution of the equation
\begin{equation}
\mathrm{vol}\left(  Q_{N}\cap\left(  \mathbb{R}_{+}^{d+1}\,\backslash
\,\lambda\left(  V,N\right)  K_{\eta}^{>}\right)  \right)  =V. \label{91}%
\end{equation}

\begin{theorem}
The unique solution to the variational problem (\ref{04}) is the surface
\begin{equation}
G_{\eta,N,V}=Q_{N}\cap\lambda\left(  V,N\right)  G_{\eta}, \label{90}%
\end{equation}
provided $N$ is large enough.
\end{theorem}

\begin{remark}
The analog of the variational problem (\ref{04}), which is obtained by the
removal of the constraint (\ref{07}), is ill posed.
\end{remark}

Note that the set $K_{\eta}^{>}$ is a convex region in $\mathbb{R}^{d+1}.$ If
at some boundary point the region $K_{\eta}^{>}$ has a unique support plane,
then this plane is of the form
\[
L_{\eta}\left(  \mathbf{n}\right)  =\left\{  \mathbf{x}\in\mathbb{R}%
^{d+1}:\left(  \mathbf{x},\mathbf{n}\right)  =\eta\left(  \mathbf{n}\right)
\right\}
\]
for the corresponding $\mathbf{n.}$ Therefore for $N$ large we have
\begin{equation}
\mathrm{vol}\left(  G_{\eta}\right)  =\frac{\mathcal{V}_{\eta}\left(  G_{\eta
}\right)  }{d+1}. \label{aa}%
\end{equation}

The proof of the above theorem relies on the following construction, which
corresponds to the variational problem (\ref{04}) a certain Wulff variational
problem. Once the correspondence is established, the proof follows easily from
the known properties of the Wulff problem.

\section{The corresponding Wulff problem.}

First we formulate the general Wulff problem. We restrict ourselves to the
symmetric case, in order to make the relation with the above maximizing
problem more transparent. Let the real function $\tau$ on $S^{d}$ be given. We
suppose that the function is continuous, positive: $\tau\left(  \mathbf{\cdot
}\right)  \geq const>0,$ and symmetric with respect to the reflections in
coordinate planes: $\tau\left(  \mathbf{n}\right)  \equiv\tau\left(
n_{1},n_{2},...,n_{d+1}\right)  =\tau\left(  \pm n_{1},\pm n_{2},...,\pm
n_{d+1}\right)  $ for any choice of signs. Then for every hypersurface $M^{d}$
(possibly with a boundary), embedded in $\mathbb{R}^{d+1},$ we can define the
\textbf{Wulff functional}
\begin{equation}
\mathcal{W}_{\tau}\left(  M^{d}\right)  =\int_{M^{d}}\tau\left(
\mathbf{n}_{x}\right)  \,ds_{x}. \label{01}%
\end{equation}
Here $x\in M^{d}$ is a point on the manifold $M^{d},$ and the vector
$\mathbf{n}_{x}$ is the unit vector parallel to the normal to $M^{d}$ at $x.$
We suppose that $M^{d}$ is smooth enough.

Suppose additionally that the hypersurface $M^{d}$ lies in fact in
$\mathbb{R}_{+}^{d+1},$ and separates the origin $O$ from infinity in
$\mathbb{R}_{+}^{d+1}.$ Let $N^{d+1}\subset\mathbb{R}_{+}^{d+1}$ be the union
of all finite components of $\mathbb{R}_{+}^{d+1}\,\backslash\,M^{d};$ we
denote the volume $\left|  N^{d+1}\right|  $ of $N^{d+1}$ by $\mathrm{vol}%
\left(  M^{d}\right)  ,$ and will call it \textit{the volume inside }$M^{d}.$
We denote by $D^{\prime}$ the collection of all such hypersurfaces $M^{d}%
$\textit{\ }in $\mathbb{R}_{+}^{d+1}$ with finite volumes. By $D_{\Lambda
}^{\prime}\subset D^{\prime}$ we denote the collection of all these
hypersurfaces $M^{d}$, for which the volume $\mathrm{vol}\left(  M^{d}\right)
$ inside\textit{\ }$M^{d}$ equals $\Lambda$.

The \textit{Wulff problem }consists in finding the lower bound of
$\mathcal{W}_{\tau}$ over $D_{\Lambda}^{\prime}:$%
\begin{equation}
w_{\tau}=\inf_{M\in D_{\Lambda}^{\prime}}\mathcal{W}_{\tau}\left(  M\right)  ,
\label{02}%
\end{equation}
as well as the minimizing surface(s) $W_{\tau}^{\left(  \Lambda\right)  },$
such that $\mathcal{W}_{\tau}\left(  W_{\tau}^{\left(  \Lambda\right)
}\right)  =w_{\tau},$ if it exists.

The answer is given by the following \textbf{Wulff construction}. Let
\begin{equation}
K_{\tau}^{<}=\left\{  \mathbf{x}\in\mathbb{R}^{d+1}:\forall\mathbf{n\;}\left(
\mathbf{x},\mathbf{n}\right)  \leq\,\tau\left(  \mathbf{n}\right)  \right\}
,\text{ and }W_{\tau}=\mathbb{R}_{+}^{d+1}\cap\partial K_{\tau}^{<}.
\label{42}%
\end{equation}
Note that the set $K_{\tau}^{<}$ is a symmetric convex bounded region in
$\mathbb{R}^{d+1}.$ If at some boundary point the region $K_{\tau}^{<}$ has a
unique support plane, then this plane is of the form
\begin{equation}
L_{\tau}\left(  \mathbf{n}\right)  =\left\{  \mathbf{x}\in\mathbb{R}%
^{d+1}:\left(  \mathbf{x},\mathbf{n}\right)  =\tau\left(  \mathbf{n}\right)
\right\}  \label{92}%
\end{equation}
for the corresponding $\mathbf{n.}$ Therefore
\begin{equation}
\mathrm{vol}\left(  W_{\tau}\right)  =\frac{1}{d+1}\mathcal{W}_{\tau}\left(
W_{\tau}\right)  . \label{46}%
\end{equation}

\begin{theorem}
\label{w} The variational problem (\ref{02}) has a unique solution
\[
W_{\tau}^{\left(  \Lambda\right)  }=\left(  \frac{\Lambda\left(  d+1\right)
}{\mathcal{W}_{\tau}\left(  W_{\tau}\right)  }\right)  ^{1/\left(  d+1\right)
}W_{\tau},
\]
which is a scaled version of the \textbf{Wulff shape }$W_{\tau}$, see
(\ref{42}).
\end{theorem}

\begin{remark}
The standard use of the name ``Wulff shape'' refers to the full surface
$\partial K_{\tau}^{<}$ itself. In our symmetric setting the surface $\partial
K_{\tau}^{<}$ is a union of $W_{\tau}$ and its multiple reflections in
coordinate planes.
\end{remark}

\begin{remark}
The variational problem (\ref{02}) might have no solutions if the function
$\tau$ is allowed to vanish.
\end{remark}

The paper \cite{T2} contains a simple proof that $\mathcal{W}_{\tau}\left(
W_{\tau}^{\left(  \Lambda\right)  }\right)  \leq\mathcal{W}_{\tau}\left(
M\right)  $ for every $M\in D_{\Lambda}^{\prime}.$ The uniqueness of the
minimizing surface is proven in \cite{T1}.

\bigskip

Now we are going to construct the Wulff problem, corresponding to the problem
(\ref{04}) (and its solution (\ref{90})). For this we have to specify the
function $\tau_{\eta}\left(  \mathbf{n}\right)  $ and the value $\Lambda$ of
the volume. First we put
\[
\Lambda=N^{d}-V.
\]
We define $\tau\left(  \mathbf{n}\right)  $ under assumption that the
normalization constant $\lambda\left(  V,N\right)  $ from the equation
(\ref{91}) equals to $1.$ This is done to simplify the notations. In this case
for $\mathbf{n}\in\Delta^{d}$ we put
\[
\tau_{\eta}\left(  \mathbf{n}\right)  =\mathrm{dist}\left(  A,L_{\eta}\left(
\mathbf{n}\right)  \right)  ,
\]
see (\ref{92}). For the remaining $\mathbf{n}$-s the function $\tau$ is
defined by symmetry.

\bigskip

Note that
\[
\tau_{\eta}\left(  \mathbf{n}\right)  +\eta\left(  \mathbf{n}\right)  =\left(
\mathbf{n},\overrightarrow{OA}\right)  .
\]
Therefore for $G$ in $D_{V}^{N}$ we have:
\[
\mathcal{V}_{\eta}\left(  G\right)  +\mathcal{W}_{\tau_{\eta}}\left(
G\right)  =\int_{G}\left(  \mathbf{n}_{x},\overrightarrow{OA}\right)
\,ds_{x}.
\]
Let $\Pi$ be the hyperplane orthogonal to the vector $\overrightarrow{OA},$
and $P$ denote the orthogonal projection on $\Pi.$ Then the last integral is
nothing else but the $d$-dimensional area $S\left(  P\left(  G\right)
\right)  $ of the set $P\left(  G\right)  ,$ so
\begin{equation}
\mathcal{V}_{\eta}\left(  G\right)  +\mathcal{W}_{\tau_{\eta}}\left(
G\right)  =S\left(  P\left(  G\right)  \right)  , \label{93}%
\end{equation}
and therefore
\begin{equation}
\sup_{G\in D_{V}^{N}}\mathcal{V}_{\eta}\left(  G\right)  \leq\sup_{G\in
D_{V}^{N}}S\left(  P\left(  G\right)  \right)  -\inf_{G\in D_{V}^{N}%
}\mathcal{W}_{\tau_{\eta}}\left(  G\right)  \label{94}%
\end{equation}
Denote by $G_{0}$ the surface made from $d$ faces of the cube $Q_{N}$
containing the vertex $O.$ Then clearly $S\left(  P\left(  G_{0}\right)
\right)  \geq S\left(  P\left(  G\right)  \right)  $ for every $G\in D_{V}%
^{N}.$ Since we suppose that the function $\eta$ vanishes in some neighborhood
of $\partial\Delta^{d},$ we have that $S\left(  P\left(  G_{\eta}\right)
\right)  =S\left(  P\left(  G_{0}\right)  \right)  ,$ provided $N$ is large
enough. Also, by the very definition of the function $\tau_{\eta}$ and in view
of the Theorem \ref{w}
\[
\inf_{G\in D_{V}^{N}}\mathcal{W}_{\tau_{\eta}}\left(  G\right)  =\mathcal{W}%
_{\tau_{\eta}}\left(  G_{\eta}\right)  ,
\]
so (\ref{93}) and (\ref{94}) imply that indeed
\[
\sup_{G\in D_{V}^{N}}\mathcal{V}_{\eta}\left(  G\right)  =\mathcal{V}_{\eta
}\left(  G_{\eta}\right)  .
\]

\end{document}